\documentclass[a4paper]{jpconf}
\usepackage{amsmath}
\usepackage{amssymb}
\usepackage{bm}
\usepackage{color}
\usepackage{graphicx}
\usepackage[normalem]{ulem}
\allowdisplaybreaks
\def\fig{./}

\begin{document}

\title{Improved Multi-Variable Variational Monte Carlo Method
Examined by High-Precision Calculations of One-Dimensional Hubbard Model}
\author{Ryui Kaneko, Satoshi Morita, and Masatoshi Imada}
\address{Department of Applied Physics, The University of Tokyo, 7-3-1
Hongo, Bunkyo-ku, Tokyo 113-8656, Japan}
\ead{rkaneko@solis.t.u-tokyo.ac.jp}

\begin{abstract}
We revisit the accuracy of the variational Monte Carlo (VMC) method by
taking an example of ground state properties for the one-dimensional
Hubbard model. We start from the variational wave functions with the
Gutzwiller and long-range Jastrow factor introduced by Capello {\it et
al.}  [Phys. Rev. B {\bf 72}, 085121 (2005)] and further improve it by
considering several quantum-number projections and a generalized
one-body wave function.  We find that the quantum spin projection and
total momentum projection greatly improve the accuracy of the ground
state energy within 0.5\% error, for both small and large systems at
half filling.  Besides, the momentum distribution function $n(k)$ at
quarter filling calculated up to 196 sites allows us direct estimate of
the critical exponents of the charge correlations from the power-law
behavior of $n(k)$ near the Fermi wave vector.  Estimated critical
exponents well reproduce those predicted by the Tomonaga-Luttinger
theory.
\end{abstract}

\section{Introduction}

Strongly correlated electron systems have attracted much interest over
the past decades.  Although rigorous ground states are obtained
analytically for some simple models, numerical methods offer powerful
tools for understanding more realistic systems.  There are several
numerical methods for the purpose of calculating low-energy properties
of the strongly correlated systems.  At finite temperatures, the
auxiliary-field quantum Monte Carlo (AFQMC)
method~\cite{qmc_1,qmc_2,qmc_3} gives exact ground states within the
statistical errors.  However, for the geometrically frustrated systems,
the errors become exponentially large, known as the negative sign
problem.  To overcome this difficulty, the Gaussian-basis Monte Carlo
(GBMC) method~\cite{gbmc_1,gbmc_2,gbmc_3} has been proposed as an
alternative to the AFQMC method, although the tractable range of
parameters are still under debate.  The dynamical mean-field theory
(DMFT)~\cite{dmft_1,dmft_2} can also be applied to strongly correlated
electron systems. However, when we wish to take into account sufficient
spatial correlations to represent fluctuations around broken symmetry
states, we need larger clusters beyond the present computational
accessibility.  At zero temperature, the exact diagonalization (ED)
method gives exact ground states.  This method, however, is not able to
handle large system sizes.  The density matrix renormalization group
(DMRG) method~\cite{dmrg_1} gives nearly exact ground states for
relatively larger systems, though it is mainly applied to
one-dimensional systems, or two-dimensional systems with cylindrical
boundary conditions.  The unbiased path-integral renormalization group
(PIRG) method ~\cite{pirg_1,pirg_2,pirg_3} has been applied to several
strongly correlated electron systems, with essentially accurate ground
states, even in the presence of geometrical frustrations.  However,
strong correlation regime requires a set of many basis functions for
accurate estimates, sometimes beyond the computational accessibility.
Among them, although a bias inevitably exists, the VMC
method~\cite{vmc_1} is applicable to large systems in the presence of
strong correlations and geometrical frustrations.

Accuracy of the calculated properties obtained by the VMC method relies
on the choice of the trial wave functions.  In most cases, the trial
wave functions $\left|\psi\right>$ are given by Slater determinants
$\left|D\right>$ supplemented by correlation factors $\mathcal{P}$ as
$\left|\psi\right> = \mathcal{P}\left|D\right>$.  One of the famous
example of this trial function is a ``projected
Bardeen-Cooper-Schrieffer (BCS)'' type wave function $\left|\psi\right>
= \mathcal{P}_{\mathrm{G}}^{\infty}\left|\mathrm{BCS}\right>$, where
$\mathcal{P}_{\mathrm{G}}^{\infty} = \prod_{i}
(1-n_{i\uparrow}n_{i\downarrow})$ denotes a Gutzwiller factor, which
prohibits double occupancy of the sites, and $\left|\mathrm{BCS}\right>
= \prod_{\bm{k}} (u_{\bm{k}}+v_{\bm{k}} c_{\bm{k}\uparrow}^{\dagger}
c_{-\bm{k}\downarrow}^{\dagger}) \left|0\right>$ denotes a BCS state,
which is the ground state of the BCS mean-field Hamiltonian.  This
projected BCS wave function is equivalent to the so-called ``resonating
valence bond (RVB)'' wave function, which is the superposition of all
dimer coverings~\cite{anderson_1,anderson_2}.  The projected BCS and RVB
wave functions provide highly accurate descriptions in two-dimensional
strongly correlated electron systems~\cite{liang_1,tahara_1}.

In principle, the trial wave functions based on Slater determinants
become accurate for higher dimensions since the mean-field-like
treatment becomes valid above the upper critical dimension.  Although it
seems difficult to capture the low-energy properties at lower spatial
dimensions by using such a trial wave function, long-range correlation
factors may restore the accuracy of the wave
functions~\cite{capello_1,capello_2}.  Capello {\it et al.}  have
calculated the ground state of the one-dimensional Hubbard model by
using the projected BCS type wave functions.  They have shown that the
long-range Jastrow factor is essential to describe the low-energy
properties.  They have also reproduced correct power-law behaviors of
the spin and charge correlation functions.

Although Capello {\it et al.}'s scheme appears to substantially
reproduce the low-energy properties of the one-dimensional model, the
errors in the energy become larger even for small system sizes,
typically more than $2\%$ for $L=18$ for intermediate interaction
strength~\cite{capello_2}.  This is mainly because they have used the
Slater determinants with a limited number of variational parameters.
The Slater determinants in their wave function are just a Fermi sea of
the tight binding model, or short-range BCS wave functions with nonzero
amplitudes at most up to third-nearest-neighbors.  In addition, in
practical numerical calculations, since the accessible system sizes are
limited, it is important to prepare the wave functions preserve the
correct quantum numbers within its accessible range of sizes.

In this paper, to obtain more accurate wave functions of the
low-dimensional systems for the smaller as well as for larger system
sizes, we revisit the variational calculations of the one-dimensional
Hubbard model, as a reference system.  By considering the multi-variable
Slater determinants and projections which restore the $\mathrm{SU(2)}$
spin rotational and spatially translational symmetries, we obtain more
accurate ground states of the model than the previous studies.  We also
calculate the momentum distribution function $n(k)$ up to 196 sites, the
first high-precision variational calculation of $n(k)$ to our knowledge,
and directly estimate the critical exponents of the charge
correlations. We compare them with the analysis by Kawakami {\it et
al.}~\cite{kawakami_1,kawakami_2} as well as Ogata {\it et
al.}~\cite{ogata_1}. Estimated critical exponents well reproduce those
obtained by the Tomonaga-Luttinger theory.

\section{Model and method}

We consider the one-dimensional Hubbard model.  The Hamiltonian is given
by
\begin{equation}
 H = -t \sum_{\left<i,j\right>,\sigma}
 \left( c_{i\sigma}^{\dagger} c_{j\sigma} + \mathrm{h.c.} \right)
 + U\sum_{i} n_{i\uparrow}n_{i\downarrow},
\end{equation}
where $c_{i\sigma}^{\dagger}$ ($c_{i\sigma}$), $n_{i\sigma}$, $t$ and
$U$ denote the creation (annihilation) operator, the number operators,
the nearest-neighbor hopping interaction, and the on-site Coulomb
interaction, respectively.

Exact ground states of the one-dimensional Hubbard model are obtained by
Lieb and Wu~\cite{lieb_wu_1}.  At half filling, the ground state is a
metal for $U=0$, and is an insulator for nonzero positive $U$. Away from
half filling, the ground state is a metal for any $U\ge 0$.  Unlike the
normal Fermi liquid, the momentum distribution of this metal does not
show a finite jump at the Fermi wave vector. Instead, it shows power-law
decays near the Fermi wave vector. This type of the ground state, which
appears in one-dimensional electron systems, is well known as the
Tomonaga-Luttinger liquid~\cite{tomonaga_1,luttinger_1}.

To obtain accurate variational description of the ground state of the
model, we use a multi-variable variational Monte Carlo (mVMC)
method~\cite{tahara_1}.  We use the trial wave functions given by
\begin{equation}
 \left|\psi\right> =
 \mathcal{P}_{\mathrm{G}}
 \mathcal{P}_{\mathrm{J}}
 \mathcal{P}_{\mathrm{dh}}
 \mathcal{L}^{S=0}
 \mathcal{L}^{K=0}
 \left|\phi_{\mathrm{pair}}\right>,
\end{equation}
where $\left|\phi_{\mathrm{pair}}\right>$, $\mathcal{P}_{\mathrm{G}}$,
$\mathcal{P}_{\mathrm{J}}$, $\mathcal{P}_{\mathrm{dh}}$,
$\mathcal{L}^{S=0}$, and $\mathcal{L}^{K=0}$ denote the one-body part,
the Gutzwiller factor, the Jastrow factor, the doublon-holon factor, the
spin quantum-number projection on to $S=0$ space, and the total momentum
projection on to $K=0$ space, respectively, as we detail later.  We
calculate the ground states of the model under the periodic boundary
conditions at half filling ($n=1$) for $N_{\mathrm{s}}=4n+2$ sites, and
at quarter filling ($n=0.5$) for $N_{\mathrm{s}}=8n+4$ sites, so that
the systems obey the closed shell conditions.

For the one-body part $\left|\phi_{\mathrm{pair}}\right>$, we use a
generalized pairing wave function defined as
\begin{equation}
 \left|\phi_{\mathrm{pair}}\right>
 = \left(
 \sum_{i,j=1}^{N_{\mathrm{s}}} f_{ij}
 c_{i\uparrow}^{\dagger}
 c_{j\downarrow}^{\dagger}
 \right)^{N_{\mathrm{e}}/2}
 \left|0\right>,
\end{equation}
where $f_{ij}$, $N_{\mathrm{s}}$, and $N_{\mathrm{e}}$ denote the
variational parameters, the number of sites, and the number of
electrons, respectively.  When all $f_{ij}$'s are optimized
simultaneously and independently, since the total number of $f_{ij}$ is
$N_{\mathrm{s}}^2$, the computation time for the optimization is highly
demanding.  To reduce the computation time to
$\mathcal{O}(N_{\mathrm{s}})$, we assume $f_{ij}$ to have a sublattice
structure, namely $f_{ij}=f_{\sigma(i)}(R_i-R_j)$, where $\sigma(i)$
denotes a sublattice index at site $i$.  Since optimized ground states
by using $f_{ij}$ with reflectional symmetries, namely $f_{ij}=f_{ji}$,
have higher energies, we do not impose reflectional symmetries on
$f_{ij}$.  Before calculating the ground states of larger system sizes,
we examine the accuracy of the wave functions using several sublattice
structures of $f_{ij}$ for smaller system sizes less than about 20
sites.  We find that optimized energies for two-sublattice structure and
$N_{\mathrm{s}}$-sublattice (fully optimized) structure are nearly the
same.  This suggests that the ground states are well-described by the
wave functions with the two-sublattice structure for the one-dimensional
Hubbard model.  In the following, we show results for mainly
two-sublattice structure, and one-sublattice translationally invariant
structure.

To include correlation effects beyond the one-body approximation, we
consider the correlation factors
\begin{eqnarray}
\label{eq:Pg}
 \mathcal{P}_{\mathrm{G}}
 &=& \exp\left( -g \sum_{i=1}^{N_{\mathrm{s}}}
 n_{i\uparrow}n_{i\downarrow} \right),
\\
\label{eq:Pj}
 \mathcal{P}_{\mathrm{J}}
 &=& \exp\left( - \sum_{i<j}
 v_{ij} (n_{i\uparrow}+n_{i\downarrow})
 (n_{j\uparrow}+n_{j\downarrow}) \right),
\\
\label{eq:Pdh}
\mathrm{and}\quad
 \mathcal{P}_{\mathrm{dh}}
 &=& \exp\left( - \sum_{m=0}^{2}\sum_{l=1}^{N_\mathrm{s}}
 \alpha_{ml} \sum_{i=1}^{N_\mathrm{s}}
 \xi_{iml} \right),
\end{eqnarray}
where $g$, $v_{ij}$, and $\alpha_{ml}$ denote variational parameters of
the Gutzwiller, Jastrow, and the doublon-holon correlation factors,
respectively.  We take into account long-range part of the Jastrow
factor, which is indispensable for reproducing the low-energy properties
of the one-dimensional Hubbard model~\cite{capello_1,capello_2}, and the
doublon-holon factor.  A variable $\xi_{iml}$ in Eq.~(\ref{eq:Pdh}) is a
many-body operator, which gives one if a doublon (holon) exists at the
$i$-th site and $m$ holons (doublons) exist at $l$-th nearest neighbors,
and gives zero otherwise.  We assume the Jastrow factor to have
translational symmetry, namely $v_{ij}=v(r_{ij})$, where
$r_{ij}=|R_i-R_j|$ is the distance between the sites $i$ and $j$.  We
note that the Gutzwiller factor and the doublon-holon factor have the
translational symmetry by definition.

In addition to the one-body part and the correlation factors, we
introduce a projection $\mathcal{L}^{S=0}$, which restores the
$\mathrm{SU(2)}$ spin rotational symmetry, and a projection
$\mathcal{L}^{K=0}$, which restores the translational symmetry. The spin
quantum-number and total momentum projections are switched on at all
optimization steps.  Since the correlation factors have the
$\mathrm{SU(2)}$ and translational symmetries, the quantum-number
projections commute with the correlation factors.

To optimize a large number of these variational parameters, we use the
stochastic reconfiguration (SR) method developed by Sorella~\cite{sr_1}.
We first optimize variational parameters $f_{ij}$ for typically $1000$
SR steps with fixed correlation factors.  After confirming the energy
convergence, we simultaneously optimize all the variational parameters
($g$, $v_{ij}$, $\alpha_{ml}$, and $f_{ij}$) for more than $4000$ SR
steps.

\section{Results}

\begin{figure}[h]
\centering%
\includegraphics[width=.46\textwidth]{\fig/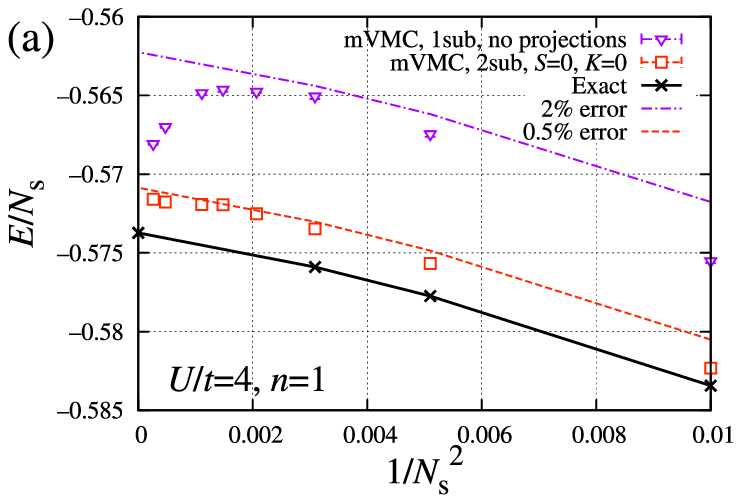}%
\hfil%
\includegraphics[width=.44\textwidth]{\fig/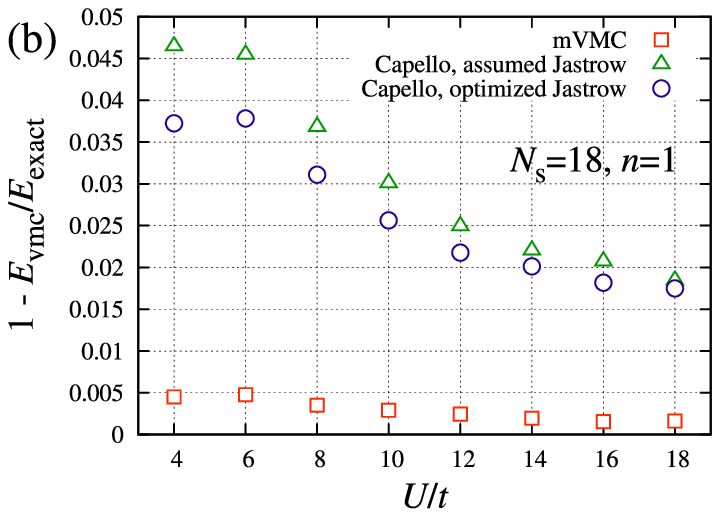}%
\caption{
Calculated energy of Hubbard model at half filling.  Statistical errors
are smaller than the symbol sizes.  (a) System size $N_{\mathrm{s}}$
dependence at $U/t=4$.  The black crosses denote the exact energies
obtained by the ED method for finite sizes and analytically at the
thermodynamic limit.  The purple down-pointing triangles and red squares
denote calculated energies by the mVMC method with the translationally
invariant and two-sublattice $f_{ij}$ trial wave functions,
respectively.  The former is obtained without the quantum-number
projections and the latter is with the spin and momentum projections to
the total singlet and zero total momentum. Two and 0.5 percent errors
measured from the exact results are plotted by dash-dotted purple and
dashed red curves, respectively.  (b) Coulomb interaction $U/t$
dependence of errors in calculated energies at half filling for
$N_{\mathrm{s}}=18$.  The red squares denote our mVMC results obtained
by using the two-sublattice $f_{ij}$ trial wave functions.  The errors
in the energy are much smaller than their results.  The green triangles
and blue circles denote the errors in the energy obtained by Capello
{\it et al.}  by using short-range BCS wave functions with the
third-nearest-neighbor amplitudes~\cite{capello_2}.  The former is
obtained with the assumed Jastrow factor and the latter is with the
optimized Jastrow factor.
}
\label{fig:ene}
\end{figure}

We obtain the ground-state energy of the one-dimensional Hubbard model
at half filling, as shown in Fig.~\ref{fig:ene}.  For the case of the
two-sublattice $f_{ij}$, the errors in the energy are within
approximately $0.5\%$, much smaller than the previous calculations by
Capello {\it et al.}~\cite{capello_2} The extended one-body part and
quantum projections improve the accuracy of the ground state energy,
even for the small system sizes.  We show results of the two-sublattice
$f_{ij}$ condition, hereafter.

\begin{figure}[h]
\centering%
\includegraphics[width=.45\textwidth]{\fig/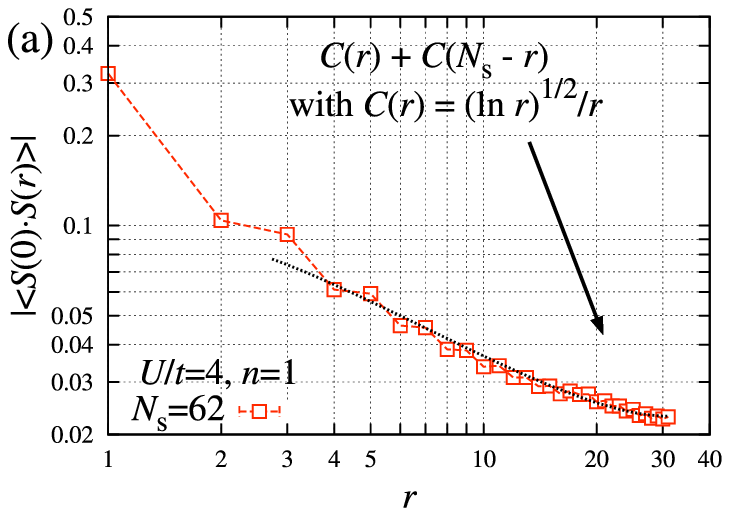}%
\hfil%
\includegraphics[width=.45\textwidth]{\fig/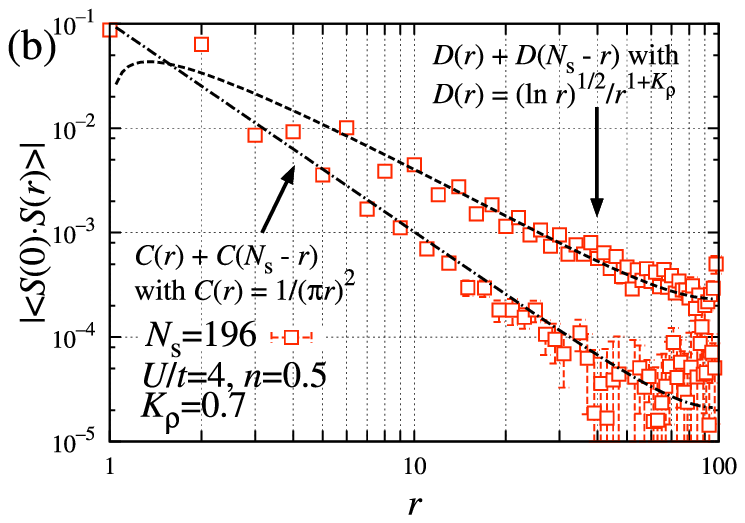}%
\caption{
Calculated spin correlation functions of Hubbard model at $U/t=4$, (a)
at half filling and (b) at quarter filling, respectively.  The red
squares denote the mVMC results.  These points are well described by the
exact asymptotic expressions of the spin correlation functions, given by
the dotted black lines.
}
\label{fig:s0sr}
\end{figure}

In the gapless one-dimensional systems, spin and charge correlations
show power-law decays~\cite{giamarchi_1,schulz_1}.  The asymptotic
expressions of the spin correlation functions are given as
\begin{equation}
\label{eq:C(r)}
 C(r)
 = \left<\bm{S}(0)\cdot\bm{S}(r)\right>
 = -\frac{1}{\pi^2 r^2} 
 + A \cos(2k_{\mathrm{F}}r) \frac{ \sqrt{\ln{r}} }{r^{K_{\rho}+1}}
 + \cdots,
\end{equation}
where $A$, $k_{\mathrm{F}}$, and $K_{\rho}$ denote a constant, the Fermi
wave vector, and a critical exponent, respectively.  The Fermi wave
vector $k_{\mathrm{F}}$ is written in terms of the filling $n$ through
the equation $k_{\mathrm{F}}=n\pi/2$.  The critical exponent $K_{\rho}$
is a function of the Coulomb interaction $U$ and the filling $n$.  At
half filling, the charge degrees of freedom are gapped out with the
exponential decay of correlations in distance, leaving only the gapless
spin correlation in the form of Eq.~(\ref{eq:C(r)}) formally with
$K_{\rho}=0$ for nonzero repulsive $U$.  Namely, the spin correlation
functions decay as $C(r)\sim A (-1)^r\sqrt{\ln{r}}
/r$~\cite{giamarchi_1,schulz_1,hallberg_1}.  Away from half filling, the
critical exponent satisfies $0<K_{\rho}<1$ for nonzero repulsive
$U$. Thus the spin correlation functions decay as Eq.(\ref{eq:C(r)}).
To include finite size effects under the periodic boundary conditions,
we compare the calculated correlation functions with $C(r) +
C(N_{\mathrm{s}}-r)$ instead of the original correlation function $C(r)$
because of the reflection effects from the boundary.  As shown in
Fig.~\ref{fig:s0sr}, calculated spin correlation functions of the
Hubbard model reproduce $\sqrt{\ln r}/r$ decay at half filling, and
$\sqrt{\ln r}/r^{1+K_{\rho}}$ and $1/r^2$ decays at quarter filling with
the critical exponent $K_{\rho}\approx 0.7$ at $U/t=4$, respectively.
This critical exponent is nearly the same as the exact value,
$K_{\rho}=0.711$, predicted by the Tomonaga-Luttinger
theory~\cite{capello_1,schulz_1,kawakami_1,kawakami_2}.

\begin{figure}[h]
\centering%
\includegraphics[width=.45\textwidth]{\fig/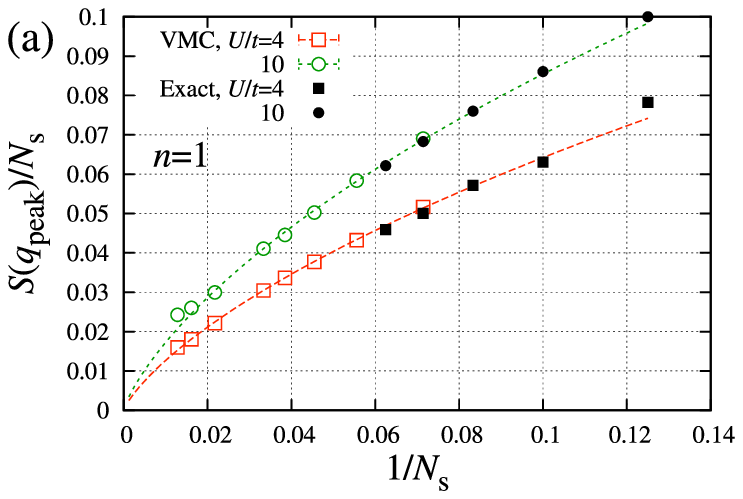}%
\hfil%
\includegraphics[width=.45\textwidth]{\fig/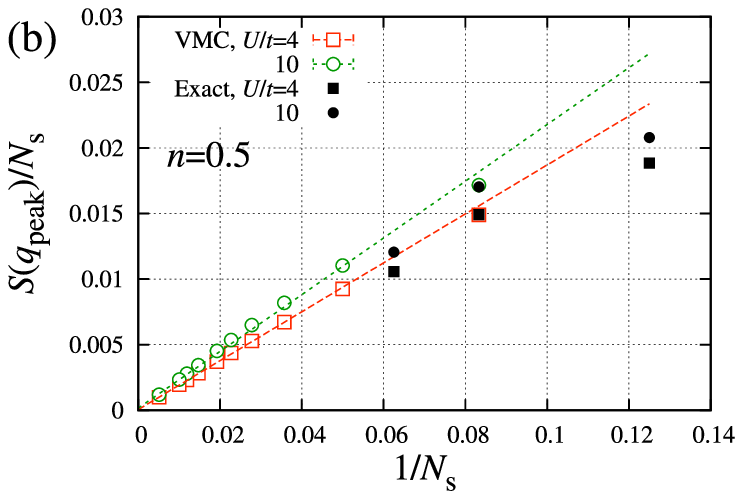}%
\caption{
Peak values of spin structure factors of Hubbard model at $U/t=4$ and
$U/t=10$ for each system size, (a) at half filling and (b) at quarter
filling.  Statistical errors are smaller than the symbol sizes.  At half
filling, the spin structure factor shows a peak at
$q_{\mathrm{peak}}=\pi$, while at quarter filling, it shows a peak at
$q_{\mathrm{peak}}=\pi/2$.  The filled symbols denote results obtained
by the ED method for the model under the periodic boundary conditions,
while the open symbols denote the mVMC results.  The red squares denote
results at $U/t=4$, and the green circles denote results at $U/t=10$.
The squared magnetizations, $S(q_{\mathrm{peak}})/N_{\mathrm{s}}$
converge to zero in the thermodynamic limit.
}
\label{fig:sqpeak}
\end{figure}

In our calculations, the one-body part of the trial wave function
includes long-range part in $f_{ij}$ and $v_{ij}$.  Although there
should be no long-range order for the one-dimensional Hubbard model, our
trial wave function could fallaciously stabilize the insulator with the
antiferromagnetic long-range order.  To examine whether our calculations
correctly reproduce the absence of the long-range order in the model, we
calculate the peak values of the spin structure factors
\begin{equation}
 S(q) = \frac{1}{3N_\mathrm{s}} \sum_{i,j}
 e^{iq(R_i-R_j)}
 \left< \bm{S}_i\cdot\bm{S}_j \right>
\end{equation}
for each system size.  For the insulator, size dependencies of the peak
values of the spin structure factor are expected to follow the scaling
\begin{equation}
 S(q_{\mathrm{peak}})
 = \frac{1}{3} \int dr\,
 e^{iq_{\mathrm{peak}}r}
 \left< \bm{S}(0)\cdot\bm{S}(r) \right>
 \sim \int_{\Lambda}^{N_{\mathrm{s}}} dr
 \frac{\sqrt{\ln r}}{r}
 \sim (\ln N_{\mathrm{s}})^{3/2},
\end{equation}
where $\Lambda$ denotes the cutoff, while for the metal, since
$1/r^{1+K_{\rho}} \gg 1/r^2$ for $r\gg 1$, we obtain
\begin{equation}
 S(q_{\mathrm{peak}})
 = \frac{1}{3} \int dr\,
 e^{iq_{\mathrm{peak}}r}
 \left< \bm{S}(0)\cdot\bm{S}(r) \right>
 \sim \int_{\Lambda}^{N_{\mathrm{s}}} dr
 \frac{1}{r^{1+K_{\rho}}}
 \sim \mathrm{const.}
 + \mathcal{O}(N_{\mathrm{s}}^{-K_{\rho}}).
\end{equation}
By using these equations, we extrapolate the squared magnetization, the
peak values of the spin structure factor $S(q_{\mathrm{peak}})$ divided
by the system size $N_\mathrm{s}$, in the thermodynamic limit.  As shown
in Fig.~\ref{fig:sqpeak}, the squared magnetizations,
$S(q_{\mathrm{peak}})/N_{\mathrm{s}}$ converge to zero in the
thermodynamic limit. These results suggest that our trial wave functions
with the long-range $f_{ij}$ and symmetry-restoring projections
$\mathcal{L}$ are able to reproduce ground states with no long-range
magnetic order with power-law decay of correlations.

\begin{figure}[h]
\centering%
\includegraphics[width=.51\textwidth]{\fig/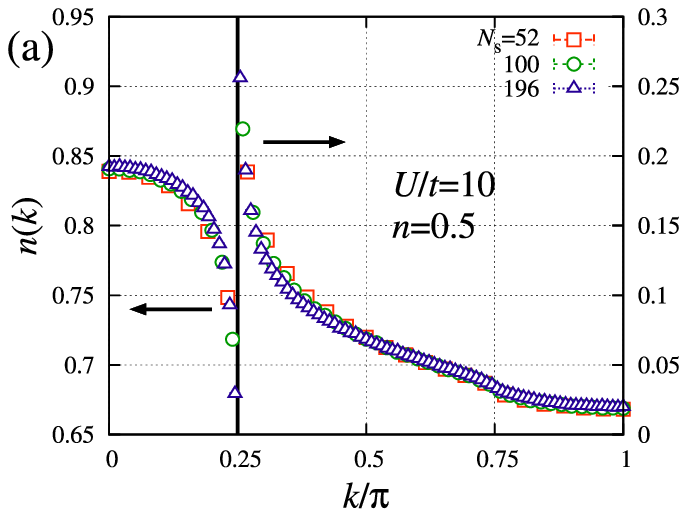}%
\hfil%
\includegraphics[width=.39\textwidth]{\fig/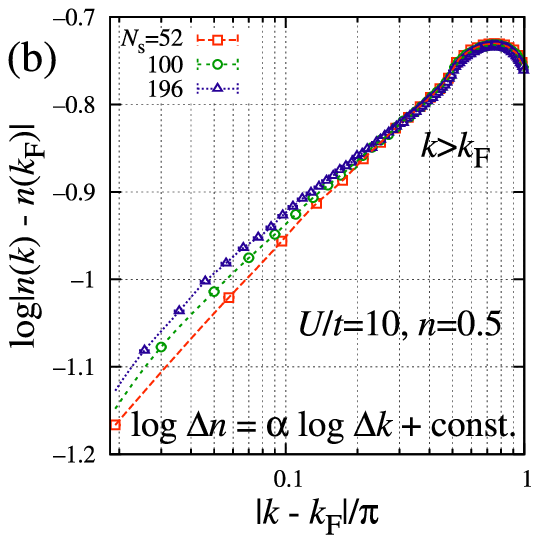}%
\caption{
(a) Momentum distribution of Hubbard model at $U/t=10$ for each system
size, at quarter filling. Power-law singularities appear at the momenta
$k_{\mathrm{F}}=\pi/4$ and $3k_{\mathrm{F}}=3\pi/4$.  (b) Logarithmic
plot of momentum distribution near $k_{\mathrm{F}}$ for each system
size.  Statistical errors are smaller than the symbol sizes.
}
\label{fig:nk}
\end{figure}

For the one-dimensional Hubbard model, the momentum distribution
\begin{equation}
 \label{eq:nk}
 n(k) = \frac{1}{2N_{\mathrm{s}}} \sum_{i,j,\sigma}
 e^{ik(R_{i}-R_{j})} \left<
 c_{i\sigma}^{\dagger} c_{j\sigma}
 \right>
\end{equation}
shows a power-law singularity
\begin{equation}
\label{eq:def_alpha}
 |n(k)-n(k_{\mathrm{F}})|\propto|k-k_{\mathrm{F}}|^{\alpha}
\end{equation}
with a critical exponent $\alpha$ near the Fermi wave vector
$k_{\mathrm{F}}$, contrary to the ordinary Fermi liquid, which shows a
jump at the Fermi wave vector~\cite{kawakami_1,kawakami_2}.  The
critical exponent $\alpha$ is written in terms of $K_{\rho}$ through the
equation $\alpha = (K_{\rho}+1/K_{\rho}-2) /
4$~\cite{parola_sorella_1,parola_sorella_2,schulz_1}.  For the infinite
Coulomb interactions, Ogata {\it et al.} have calculated the momentum
distribution $n(k)$ of the metal ground states by using the Bethe-ansatz
wave functions, and have estimated the critical exponents $\alpha$ from
Eq.~(\ref{eq:def_alpha})~\cite{ogata_1}.  We calculate the momentum
distribution of the Hubbard model at quarter filling up to 196 sites.
As shown in Fig.~\ref{fig:nk}(a), calculated momentum distributions show
power-law behavior near the wave vector $k_{\mathrm{F}}$.  Another
singularity also appears at $k=3k_{\mathrm{F}}$, as expected in the
previous studies~\cite{ogata_1}.

\begin{figure}[h]
\centering%
\includegraphics[width=.45\textwidth]{\fig/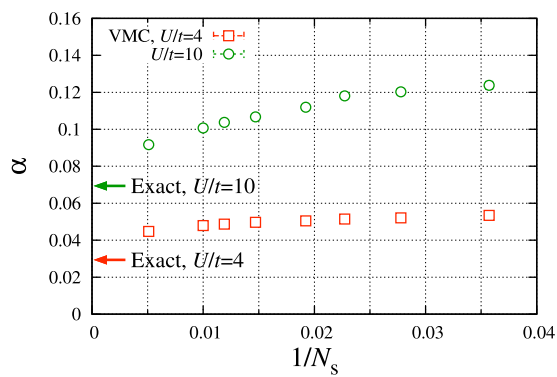}%
\caption{
Estimated critical exponent $\alpha$ from relation
$|n(k)-n(k_{\mathrm{F}})|\propto|k-k_{\mathrm{F}}|^{\alpha}$ near
$k_{\mathrm{F}}$.  Statistical errors are smaller than the symbol sizes.
The red squares denote results at $U/t=4$, and the green circles denote
results at $U/t=10$.  These values are nearly the same as the exact
results, $\alpha=0.029$ ($U/t=4$) and $\alpha=0.069$ ($U/t=10$),
predicted by the Tomonaga-Luttinger
theory~\cite{capello_1,schulz_1,kawakami_1,kawakami_2}.
}
\label{fig:alpha}
\end{figure}

To estimate the critical exponent $\alpha$, we plot $\ln
|n(k)-n(k_{\mathrm{F}})|$ as a function of $|k-k_{\mathrm{F}}|$ for each
system size, as shown in Fig.~\ref{fig:nk}(b), and extrapolate the
slopes from data points by using Eq.~(\ref{eq:def_alpha}).  Since large
finite-size effects appear near $|k-k_{\mathrm{F}}|/\pi \lesssim 0.1$,
we use data that satisfy $|k-k_{\mathrm{F}}|/\pi \in [0.1,0.4]$ for
fitting.  As shown in Fig.~\ref{fig:alpha}, as the system size
$N_{\mathrm{s}}$ goes to infinity, the estimated critical exponent
$\alpha(N_{\mathrm{s}})$ gradually converges to the exact value,
$\alpha=0.029$ at $U/t=4$ and $\alpha=0.069$ at $U/t=10$, predicted by
the Tomonaga-Luttinger
theory~\cite{capello_1,schulz_1,kawakami_1,kawakami_2}.

\section{Summary}

To improve the VMC method, we revisit the variational description of the
ground states of the one-dimensional Hubbard model, by considering the
multi-variable Slater determinants as well as the long-range correlation
factors.  We find that the quantum spin projection and total momentum
projection greatly improve the accuracy of the ground state energy with
less than $0.5\%$ error, irrespective of the system size.  By
calculating the momentum distribution function $n(k)$ of the model at
quarter filling up to 196 sites, we have shown that our method is able
to reproduce power-law decays at the momenta $k_{\mathrm{F}}$ and
$3k_{\mathrm{F}}$.  By directly estimating the critical exponents of the
charge correlations from the power-law behavior of $n(k)$ near
$k_{\mathrm{F}}$, we have also shown that estimated critical exponents
well reproduce those predicted by the Tomonaga-Luttinger theory.

\ack

The mVMC codes used for the present computation are based on that first
developed by Daisuke Tahara.  They thank Takahiro Misawa for fruitful
discussions.  This work was supported by a Grant-in-Aid for Scientific
Research (Grant Nos. 22104010 and 22340090), from MEXT, Japan. A part of
this research has been funded by the Strategic Programs for Innovative
Research (SPIRE), MEXT, and the Computational Materials Science
Initiative (CMSI), Japan.

\section*{References}

\end{document}